\documentclass{article}

\usepackage{arxiv}

\usepackage[utf8]{inputenc} 
\usepackage[T1]{fontenc}    
\usepackage{hyperref}       
\usepackage{url}            
\usepackage{booktabs}       
\usepackage{amsfonts}       
\usepackage{nicefrac}       
\usepackage{microtype}      
\usepackage{lipsum}

\usepackage{amsmath,amssymb}
\usepackage{bm}

\usepackage{graphics}

\usepackage{graphicx}

\title{Cosmological inflation driven by a scalar torsion function}

\author{
  T. M. Guimar\~aes\\
  Instituto Federal de Educa\c{c}\~ao, Ci\^encia e Tecnologia do Paran\'a (IFPR),\\ Ivaipor\~a, PR, 86870-000, Brazil\\
  \texttt{thiago.moreira@ifpr.edu.br} \\
   \And
 R. de C. Lima\\
  Universidade Estadual Paulista (UNESP), Faculdade de Engenharia de Guaratinguet\'a, Departamento de F\'isica e Qu\'imica,\\ Guaratinguet\'a, SP, 12516-410, Brazil\\
  \texttt{rc.lima@unesp.br} \\
  \And
 S. H. Pereira\\
  Universidade Estadual Paulista (UNESP), Faculdade de Engenharia de Guaratinguet\'a, Departamento de F\'isica e Qu\'imica,\\ Guaratinguet\'a, SP, 12516-410, Brazil\\
  \texttt{s.pereira@unesp.br}
}

\begin{document}
\maketitle

\begin{abstract}
A viable model for inflation driven by a torsion function in a Friedmann background is presented. The scalar spectral index in the interval $0.92\lesssim n_{s}\lesssim 0.97$ is obtained in order to satisfy the initial conditions for inflation. The post inflationary phase is also studied, and the analytical  solutions obtained for scale factor and energy density generalizes that ones for a matter dominated universe, indicating just a small deviation from the standard model evolution. The same kind of torsion function used also describes satisfactorily the recent acceleration of the universe, which could indicate a possible unification of different phases, apart form specific constants.
\end{abstract}

\keywords{Inflation; torsion; scalar and tensor perturbations.}

\section{Introduction}\label{sec:introd}

In Quantum Field Theory, the irreducible and unitary representations of the Poincaré group corresponds to stable particles; these representations are given by mass and spin. More precisely, Poincaré-Lie algebra can be classified by two values of the Casimir operator, which are the square of the mass, $m^{2}$, and the square of the angular momentum, $S^{2} = s (s + 1)$, where $s$ is the particle spin \cite{Wald:1984rg}. Thus, it is natural to think that a generalized theory of gravitation takes into account not only the contribution of mass, but also the spin as a source of the gravitational field \cite{Hehl:1976kj}. In General Relativity, the space-time $\mathcal{M}$ is curved, that is, the mathematical structure of the group that acts on an orthonormal base, in the space tangent to $\mathcal{M}$, is the Lorentz group, when the torsion is introduced and related to the spin $s$, a theory of gravitation is created in a scenario which the Poincaré group acts on the basis of the affine space in the tangent space of $\mathcal{M}$. 

It is well known that the presence of fermionic matter in spacetime, with its intrinsic spin angular momentum, is responsible for torsion effects in the manifold of interest. A direct effect produced by torsion is the inclusion of an asymmetric term to the affine connection and, consequently, increasing the degrees of freedom of the physical system. From a cosmological point of view, this implies a more complete scenarios to understand the dynamics and structure between matter and gravity. As a direct example of physical interest, one can quote condensate of fermionic particles as a spin fluid \cite{Hehl:1974cn,Sabbata90,Sabbata94}, and other case is explored for scenarios with an effective ultraviolet cutoff in quantum field theory for fermions \cite{Poplawski:2009su}. However, it is worth mentioning that there are also studies on the theoretical constructions about the nature of torsion with the inclusion of matter in spacetime that date from the 60s-70s, through of the ECKS gravitational theory \cite{Sciama:1964wt,Hehl:1971qi,Hehl:1976kj}, which describes invariance of local gauge in relation to the group of Poincar\`e, well reviewed in \cite{Shapiro:2001rz}.  

For a more interesting study concerning the evolution of the universe, one of the first derivation of Friedmann-Robertson-Walker (FRW) equations with torsion was presented in \cite{tsamPRD} and very recently the analyzes of classical Einstein-Cartan gravity for cosmological spacetimes with nonzero torsion was rediscussed in \cite{Pasmat2017,Kranas2019,Barrow2019,Pereira:2019yhu}. The FRW cosmological scenarios are great environment to analyze the consequences of torsion effects due to high symmetry in its metric, which preserves the symmetry associated to Ricci curvature tensor and makes the Einstein tensor and energy-momentum tensor to preserve their symmetric form. 

A simple and direct approach to understand the behavior of a matter field endowed with torsion effects under the evolution of a FRW background is introducing a single scalar function representing the torsion field, depending only on time, as established by \cite{Capozziello:2008kb,Poplawski:2010kb,Olmo:2011uz,Jimenez:2015fva}. Currently, Kranas et al. \cite{Kranas2019} showed how torsion can change considerably the standard evolution of the Friedmann models, in addition to checking the conditions of its influence on abundance of helium-4 in the early universe. A Friedmann-like universe with weak torsion was studied in \cite{Barrow2019} in a dynamical system approach. Motivated by these latest works, Pereira et al. \cite{Pereira:2019yhu} have assumed more general expressions to torsion function, some of them proportional to matter density and other free parameters, and analyzed the constraints of the free parameters with current observational data in order to reproduce the current accelerated expansion of the universe. Even more recently, the works \cite{Marques:2019ifg, Bose:2020mdm, Bolejko:2020nbw, Cruz:2020hkh} explored low-redshift constraints by weak torsion field, modified torsion field, distinction between torsion and the dark sector, and late time cosmology by torsion effects, respectively, all in FRW cosmological scenarios. Although being difficult to obtain experimental evidence about the real effects of space-time torsion, once that such effects require a high energy density at cosmological events, there are predictive studies of how torsion signatures could manifest via inflation \cite{Gonzalez-Espinoza:2019ajd, Gasperini:1986mv} and gravitational waves \cite{Valdivia:2017sat,Izaurieta:2019dix}. In the works \cite{March:2011ry,Hehl:2013qga} there are interesting studies and suggestions for experimental tests of non-zero torsion for gravity, concerning the Moon and Mercury motion.  

In the present paper we extend the analysis based on the specific torsion function given by Eq. (\ref{phimn}), previously studied in \cite{Pereira:2019yhu} and which is totally compatible with recent accelerated universe, in order to describe also the inflationary period, in addition to an analysis of scalar and tensor perturbation theory and determination of the spectral index $n_s$.

The paper is organised as follows. In Section 2 we present the main equations of a Friedmann cosmology with torsion. In Section 3 we study the inflationary phase driven by a specific torsion function. In Section 4 the solutions for pre and post-inflationary period is presented, and in Section 5 the scalar and tensor perturbations with torsion are obtained. Main conclusions are in Section 6.

\section{FRW cosmology with torsion}
\label{sec:torsion}
We started this section with a quick recap of the treatment of classical gravitational dynamics in the presence of torsion in a FRW background, based on the references \cite{Kranas2019, Pereira:2019yhu}. The Einstein equation, current theory of greater accuracy in the description of gravity, is 
\begin{equation}\label{1}
 G_{\mu\nu} = \kappa T_{\mu\nu},    
\end{equation}
where $G_{\mu\nu} = R_{\mu\nu} - \frac{1}{2}Rg_{\mu\nu}$ is the Einstein tensor, described in terms of the Ricci tensor and its scalar curvature. The constant ${\kappa = 8\pi G}$ complies a dimension relationship between geometry and the content of matter (or energy) represented by the energy momentum tensor on the right hand side of (\ref{1}).

The curvature, containing geometric information, is implicit in the Ricci tensor ${R_{\mu\nu}(\Gamma^\alpha_{~~~\mu\nu})}$, which in turn brings the information of the affine connection of spacetime. In the situation where torsion is present, the affine connection becomes an antisymmetric object and is rewritten as $\Gamma^\alpha_{~~~\mu\nu}=\Tilde{\Gamma}^\alpha_{~~~\mu\nu}+K^\alpha_{~~~\mu\nu}$, where $\Tilde{\Gamma}^\alpha_{~~~\mu\nu}$ represents the symmetric connection part  and $K^\alpha_{~~~\mu\nu}$ defines the contorsion tensor, decomposed into
    \begin{equation}\label{2}
        K^\alpha_{~~~\mu\nu} = S^\alpha_{~~~\mu\nu} + S_{\mu\nu}^{~~~\alpha} + S_{\nu\mu}^{~~~\alpha},
    \end{equation} 
where $S^\alpha_{~~~\mu\nu}$ are torsion tensor totally antisymmetric in its covariant indices, $S^\alpha_{~~~\mu\nu} = -S^\alpha_{~~~\nu\mu}$. This object characterize a direct coupling with the energy-momentum  tensor according to Cartan's field equation, $S_{\alpha\mu\nu} = -\frac{1}{4}\kappa (2s_{\mu\nu\alpha}+g_{\nu\alpha}s_\mu - g_{\alpha\mu}s_\nu)$, being the terms $s_{\alpha\mu\nu}$  and $s_\alpha=s^\mu_{~~\alpha\mu}$ the tensor and vector spin of matter, respectively. Since torsion can be physically understood as a link between spacetime and the intrinsic angular momentum of matter content  \cite{Pasmat2017}, the equation \eqref{1} is treated as an Einstein-Cartan equation of gravitation through the use of the contorsion \eqref{2} in the connection. 

In the light of \cite{Kranas2019}, for the scenario of a homogeneous and isotropic Friedmann background, the torsion tensor and the associated vector are
\begin{equation}\label{3}
     S_{\alpha\mu\nu} = \phi (h_{\alpha\mu} u_\nu - h_{\alpha\nu}u_\mu)   \hspace{1cm},\quad  S_\alpha = -3\phi u_\alpha, 
    \end{equation}
where $\phi=\phi(t)$ is an unique time dependent function representing torsion contribution due to homogeneity of space in FRW background, $h_{\mu\nu}$ is a projection tensor, symmetric and orthogonal to the 4-vector velocity $u_\mu$. Based on this premise and applying the Einstein-Cartan equations with the terms $S_{\alpha\mu\nu}(\phi(t))$ and $S_\alpha(\phi(t))$ introducing the torsion effect, under the FRW metric, the Friedmann equations and its continuity equation \cite{Kranas2019} are  
\begin{equation}\label{4}
        \bigg(\frac{\dot{a}}{a}\bigg)^2 =\frac{8\pi G}{3}\rho - \frac{k}{a^2}-4\phi^2 - 4\bigg(\frac{\dot{a}}{a}\bigg)\phi,
    \end{equation}
    \begin{equation}\label{5}
         \frac{\ddot{a}}{a} =-\frac{4\pi G}{3}(\rho+3p) - 2\dot{\phi} - 2\bigg(\frac{\dot{a}}{a}\bigg)\phi,
    \end{equation}
    \begin{equation}\label{6}
        \dot{\rho}+3(1+\omega)H\rho + 2(1+3\omega)\phi \rho = 0,
    \end{equation}
where ${k}$, ${\rho}$ and ${p}$ are the curvature parameter, energy density and pressure of matter, respectively. We also assume a barotropic matter density, satisfying ${p = \omega\rho}$. Given a $\phi(t)$ function, which contain all torsion information of the system, the solution of the above system of equations will furnish the complete evolution of the universe.

Some solutions with a constant torsion function where studied in \cite{Kranas2019}, however a torsion function dependent on the matter energy density is a much more realistic choice. In this regard, we cite a more general proposal to describe the temporal function ${\phi(t)}$, in terms of the energy density of matter ${\rho_m(t)}$ and the Hubble constant $H(t)$, as
\begin{equation}
\phi(t) = -\alpha H_{0} \bigg(\frac{H_{0}}{H(t)}\bigg)^{m}\bigg(\frac{\rho_m(t)}{\rho_{0c}}\bigg)^{n},\label{phimn}
\end{equation}
where $\alpha$ is a dimensionless constant, $\rho_{0c}$ and $H_{0}$ are the critical energy density and present Hubble constant, respectively. This model was very well explored in \cite{Pereira:2019yhu} and the constraints of the free parameters $\alpha$, $m$ and $n$ to drive the recent acceleration of the universe where obtained based on observational data of Supernovae type Ia and $H(z)$ data. The range of values for the parameters $(m,n)$ were $(-3.6^{+6.9}_{-7.6},$ $-1.5^{+2.4}_{-2.7})$, with $\alpha=0.30^{+0.51}_{-0.31}$. The matter density parameter and Hubble parameter were $\Omega_m=0.62^{+0.39}_{-0.62}$ and $H_{0}=68.7\pm 2.2$, respectively. Such model describes quite well the effect of a late time acceleration of the universe, where the dark energy effects originates from the torsion of the space-time.

In which follows, we will present a cosmological model driven by a single torsion function, \eqref{phimn}, assuming $m=n\approx 1$, in order to obtain analytical solutions and maintain some similarity with \cite{Pereira:2019yhu}, once these values are allowed for that model. Also, we will assume the curvature parameter $k = 0$, for simplicity, in the first Friedmann equation \eqref{4}, thereby \eqref{4} and \eqref{phimn} becomes 
\begin{align}
H(t)^{2} &= \frac{1}{3}\kappa \rho(t)-4\phi^{2}-4H(t)\phi(t),\label{eqH}\\
\phi(t) &= -\alpha H_{0} \left(\frac{H_{0}}{H(t)}\right)\left(\frac{\rho(t)}{\rho_{0}}\right),\label{phi}
\end{align}

\noindent where $H(t)\equiv \dot{a}/a$ and $\rho(t)$ is a time dependent energy density. Note that the equation \eqref{phi} is very attached to $H_{0}$ and $\rho_{0}$ constants, thus, we can construct a more general form for $\phi(t)$. Using \eqref{eqH} and \eqref{phi}, and fixing the time $t_{*}$, related to some specific cosmological era, we have

\begin{equation}\label{Hp}
\frac{H^{2}_{0}}{\rho_{0}}= \left[\frac{\kappa}{3(1-2\alpha)^2} \right],
\end{equation}

\noindent where $H(t_*)\equiv H_{0}$ and $\rho(t_{*})\equiv \rho_{0}$. It allows us to rewrite the equation \eqref{phi} as 

\begin{equation}\label{phig}
\phi(t)=-\alpha\frac{\kappa}{3(1-2\alpha)^2}\frac{\rho(t)}{H(t)},
\end{equation}

\noindent being more compact in terms of cosmological constants. Now, we are able to discuss an inflationary scenario driven by \eqref{eqH} and \eqref{phi} and how this equations will behave in a pre and post-inflationary era.

\section{Inflation driven by torsion}
\label{sec:inflation}
Now, we will study an inflationary model based on the above torsion function, eq. \eqref{phi}, assuming a constant energy density specific to inflationary phase $\rho_0$, and  also the constant Hubble parameter during inflation $H_0$. 
 
The energy density of $\phi(t)$ dominates the universe during inflation because in this period we are at high energies and dark matter are not yet fully formed. A torsion-dominated early universe undergone a phase of inflationary expansion without the need of a cosmological constant, or the presence of an inflationary field \cite{Kranas2019}. This leads to the interpretation of a vacuum solution for the $\phi$ function. Therefore, assuming $\dot{\rho} \sim 0$ for inflation in the continuity equation \eqref{6} and using \eqref{phi}, we get
\begin{equation}\label{rhoinflac}
\rho \approx \rho_{0}\left[\frac{3}{2}\frac{(1+\omega)}{\alpha(1+3\omega)}\frac{H^{2}}{H^{2}_{0}}\right].
\end{equation}

\noindent The equation \eqref{rhoinflac} shows that energy density is constant during the inflationary period, $\rho(t)\sim \rho_0$ if the Hubble radius is constant during inflation, $H(t)\sim H_0$, which leads to a constraint on the $\alpha$ and $\omega$ parameters:
\begin{equation}\label{alphaomega}
\alpha = \frac{3}{2}\frac{(1+\omega)}{(1+3\omega)} \Leftrightarrow \omega = \frac{1}{3}\frac{(2\alpha-3)}{(1-2\alpha)}.
\end{equation}

\noindent Whether $\alpha < 0$ or $\alpha > 0$, these relations between $\alpha$ and $\omega$ are valid just for inflationary period, acting like a inflationary mechanism.


A description of the initial density of matter must now be analyzed for the process that will generate the conditions of inflation. Writing \eqref{eqH} as \begin{equation}\label{soloneH}
H = \pm\left[\frac{\kappa\rho}{3(1-2\alpha)^2} \right]^{1/2},
\end{equation}
and using it into \eqref{rhoinflac}, the condition for initial matter density is:
\begin{equation}
\rho_{0} = \frac{2\alpha H_{0}^{2}}{\kappa}\frac{(1+3\omega)}{(1+\omega)}(1-2\alpha)^{2}.
\end{equation}

\begin{figure}[htb]
\center{\includegraphics[width=10.0cm]
{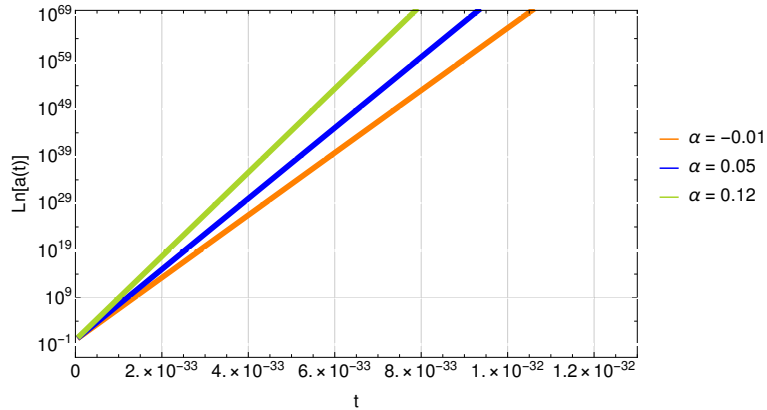}}
\caption{\label{fig:my-label} Comparison in the evolution of the scale factor ${a(t)}$ for different $\alpha$ values, whose parameters are negative and positive, in the inflationary period.}\label{Fig1}
\end{figure}

From ${H = \dot{a}/a}$ of the equation \eqref{soloneH}, with ${\rho \approx \rho_{0}}$ in the inflationary era, the solution for $a(t)$ is:
\begin{equation}
a(t) = a_{i}e^{\beta t},
\end{equation}
where $\beta \equiv \left(\frac{\kappa}{3} \frac{\rho_{0}}{(1-2\alpha)^{2}}\right)^{1/2}$. Notice that $\beta t = N$, being $N$ the e-fold number. The behavior of the scale factor during inflationary era is shown in Fig. \ref{Fig1} for different $\alpha$ values. It is possible to see that $\alpha$ parameter has strong influence during the inflationary period of the universe.

\section{Pre and post-inflationary cosmological period}
\label{sec:constraint}
Now, in this section, firstly we will explore the classical consequences of the behavior of the scale factor and density of matter in the evolution of the pre-inflationary universe. Assuming that equation \eqref{soloneH} is compactly rewritten as ${H = \gamma\sqrt{\rho}}$, where ${\gamma \equiv \pm \sqrt{\kappa/3}(1-2\alpha)^{-1}}$, using the equation \eqref{phig} in the continuity equation \eqref{6}, we get
\begin{equation}
\dot{\rho}+3(1+\omega)\gamma\rho^{3/2}-\frac{2\alpha}{\gamma \rho_{i}}(1+3\omega)\rho^{3/2}=0,
\end{equation}
where $\rho_{i}$ is the initial content of energy density to pre-inflatio-\ nary phase,  whose solution is
\begin{equation}
\rho=\frac{4\rho_{i}}{(\beta_{i}(t-t_{i})\sqrt{\rho_{i}}+2)^{2}},
\end{equation}
where ${\beta_{i} \equiv 3(1+\omega)\gamma - 2\alpha(1+3\omega)/(\gamma \rho_{i})}$.

Using again $H=\gamma \sqrt{\rho}$, from \eqref{soloneH}, the solution for scale factor is described as
\begin{equation}\label{nseq}
a(t)=\frac{a_{i}}{2}(2+\beta_{i}\sqrt{\rho_{i}}(t-t_{i}))^{\frac{2}{3(1+\omega)-2\alpha(1+3\omega)}},
\end{equation}
where $a_i$ represents the scale factor at $t=t_i$. The solution \eqref{nseq}
 shows that a strictly classical Eisntein-Cartan theory of gravitation leads to a non-singular Big Bang at $t=t_{i}$, with $\phi$ given by \eqref{phig}, this behaviour for $U(4)$ theories is well known from the literature \cite{Hehl:1974cn}. Also, it is possible to see that the above solution for the scale factor at $t>>t_i$ has a dominant behaviour as
\begin{equation}
a(t)\sim t^{\frac{2}{3(1+\omega)-2\alpha(1+3\omega)}},
\end{equation}
which reproduces the matter dominated phase in standard cosmology model if $\alpha = 0$.

The relation between $a(t)$ and $\rho(t)$ can also be verified when substituting $\rho=\frac{H^{2}}{\gamma^{2}}$ into continuity equation. Therefore, integrating \eqref{6}, we found
\begin{equation}\label{rhocosmoatual}
\rho(t) = \rho_{i}\left(\frac{a(t)}{a_{i}} \right)^{-3(1+\omega)+2\alpha(1+3\omega)},
\end{equation}
which also reproduces the matter dominated solution if $\rho_i  \longmapsto \rho_{0c}$, the current critical matter density, and $a_i = a_0$ for this same scenario of the today's universe.

\section{Scalar and Tensor perturbations with torsion}
\label{sctep}
Despite of the absent of a scalar field in inflationary mechanism studied here, the hurdle to obtain the linearized Einstein equations relays on the torsion therms. For sake of simplicity, in this first approach we will closely follow \cite{Mukhanov:1990me}, i.e., we will work on Poisson gauge, using the following  perturbed metric:

\begin{equation}\label{mp}
ds^{2}= - a^{2}(\eta)(1-2\Phi)d\eta + a^{2}(\eta)[(1+2\Psi)\delta_{ij}+2t_{ij}]dx^{i}dx^{j}, 
\end{equation}

\noindent where $\Phi$ and $\Psi$ are Bardeen potentials, which represent the scalar perturbations, and $t_{ij}$ is a tensor with no divergence and trace, representing the tensor perturbation. Henceforth, the conformal time $a(\eta)d\eta=dt$, will be assumed.\\

The perturbed metric (\ref{mp}) leads to the following components of the linearized Einstein equations:

\begin{eqnarray}
\delta G^{\eta}{}_{\eta} &=&\frac{6}{a^{2}}\bigg[\frac{\Delta \Psi }{3}-\Phi  \mathcal{H}^2-\Psi ' \mathcal{H}-4 \Phi  \phi  \mathcal{H}-4 \Phi  \phi^2 - 2 \phi\Psi '\bigg],\label{p00}\\
\delta G^{\eta}{}_{i}&=&-\delta G^{i}{}_{\eta} = \frac{2 \partial_{i}\left(\Phi  \mathcal{H}+2 \Phi\phi+\Psi\right)}{a^2},\\
\delta G^{i}{}_{j} &=& \frac{6}{a^{2}}\bigg[-\frac{\Phi  a''}{a}+\frac{\Delta ( \Psi - \Phi)}{4}+\frac{\mathcal{H}^2 \Phi }{2}-\frac{\mathcal{H} \Phi '}{2} -\mathcal{H} \Psi '-2 \mathcal{H} \Phi  \phi-\frac{\Psi ''}{2}-2 \Phi  \phi ^2-2 \Phi\phi'-\phi\Phi'\nonumber\\
&-&2 \phi \Psi' \bigg] \delta^{i}_{j} + \partial^{i}\partial_{j}(\Phi-\Psi).
\end{eqnarray}

\noindent Note that the static universe, $H=0$, ensure $\phi=0$, hence the Newtonian limit, given by (\ref{p00}), remains valid. Also, we are not using an effective stress-tensor, as in \cite{Hehl:1976kj}, instead, we are treating the torsion as part of the geometrical part of the Einstein equations. Therefore, $\delta G^{i}{}_{j} = 0$ for $i\neq j$, leads to $\Psi=\Phi$, as usual. 

Once torsion does not contribute with tensor perturbations, it has the canonical form:

\begin{equation}\label{gw}
t''_{ij}+2\mathcal{H}t'_{ij}-\Delta t_{ij} = -16 \pi G a^{2} \delta T^{i}{}_{j(T)},    
\end{equation}

\noindent with $\delta T^{i}{}_{j(T)}=0$ implying in sourceless gravitational waves, travelling in the speed of light.

Aiming on Mukhona-Sasaki equation, we can use the relation 

\begin{equation}\label{pror}
\delta p =c^{2}_{s}\delta \rho + \tau \delta S,
\end{equation}

\noindent where $c^{2}_{s}$ is the speed of sound, $\tau=\frac{\partial p}{\partial S}\left|_{\rho}\right.$, with $S$ being the entropy of the system. Using the linearized Einstein equation, (\ref{pror}) gives \footnote{For those familiar with Kranas et-all article \cite{Kranas2019}, the same equation can be obtained by perturbation of the equation (6) and (16) in ref.\cite{Kranas2019}.}

\begin{eqnarray}\label{muk1}
&\Psi ''& - c^{2}_{s}\Delta \Psi + 3(1+c^{2}_{s})\mathcal{B}\Psi'+ \Psi\left(\mathcal{B}^{2}(1+3c^{2}_{s})+2\mathcal{B}'\right) =4\pi G a^{2}\delta S,
\end{eqnarray}









\noindent where $\mathcal{B}=2\phi+\mathcal{H}$. If $\phi\rightarrow 0$ we recover the standard Mukhanov-Sasaki equation. In the present model, the speed of sound is exactly the same as canonical model:

\begin{equation}
c^{2}_{s}=\frac{dp}{d\rho}=\omega.
\end{equation}

\noindent This result shows that the torsion does not affect the speed of sound, as expected. 

As usual in perturbative models, it is possible to eliminate the $\Psi'$ term using a generic change of variable. Assuming $\Psi =u(\eta,x^{i})f(\eta)$, the equation (\ref{muk1}) turns into

\begin{eqnarray}
u''&+&\frac{f''}{f}u-c^{2}_{s}\Delta u+2u'\left(\frac{f'}{f}+\frac{3}{2}(1+c^{2}_{s})\mathcal{B}\right)+u\bigg[\frac{f''}{f}+3\frac{f'}{f}(1+c^{2}_{s})\mathcal{B}+\mathcal{B}^{2}(1+3c^{2}_{s})+2\mathcal{B}'\bigg] = \mathcal{S},
\end{eqnarray}

\noindent with $\frac{f'}{f}=-\frac{3}{2}(1+c^{2}_{s})\mathcal{B}$, $\frac{f''}{f}=\frac{3}{2}(1+c^{2}_{s})\left[\frac{3}{2}(1+c^{2}_{s})\mathcal{B}-\mathcal{B}'\right]$, and $\mathcal{S}=4\pi G f^{-1} a^{2}\delta S$, leading to

\begin{eqnarray}
u''&-&c^{2}_{s}\Delta u-\frac{u}{4}\bigg[\mathcal{B}^{2}(1+c^{2}_{s})(5+9c^{2}_{s})+2(1-c^{2}_{s})\mathcal{B}'\bigg] = \mathcal{S}.
\end{eqnarray}

\noindent The above equation can be rewrite as 

\begin{equation}\label{mk12}
u'' - c^{2}_{s}\Delta u - u\left(\frac{\theta''}{\theta}+G\right)=\mathcal{S}. 
\end{equation}

\noindent For adiabatic perturbations, $\mathcal{S}=0$, we have

\begin{equation}\label{mk2}
u'' + k^{2} u - u\left(\frac{\theta''}{\theta}+G\right)=0. 
\end{equation}

\noindent The solutions of this equation can be obtained in two cases, the first one is the long-wavelength limit, when gravity dominates and $k^{2}u$ can be neglected. The second one is short-wavelength, when $k^{2}\ll [(\theta''/ \theta)+G]$.  

\subsection{Long-wavelength limit}

In the long-wavelength limit, the equation (\ref{mk2}) takes the form 

\begin{equation}\label{ulw}
u'' - u\left(\frac{\theta''}{\theta}+G\right)=0. \end{equation}

\noindent This equation has the same form of equation ($4.7$) in \cite{Basak:2011wp}, although, the interpretation of the extra term, $G$, is very different. In the referenced work, $G$ represents the contribution of a Non Standard Spinor to Mukhanov-Sasaki equation, here, $G$ represents the contribution of torsion terms, which are objects with very distinct nature.\\

Assuming the solution  $u=u_{(can)}h$, where $u_{(can)}$ is the solution of canonical Mukhanov-Sasaki equation, and it can be recovered when the correction for torsion, $h$, tends to 1, that is $\phi\rightarrow 0$. As we previously know the solution for $u_{(can)}$, the focus of our attention must be the equation for $h$. Using the assumed solution for $u$ in (\ref{ulw}), we have   
\begin{equation}
h'' + 2\left(\frac{u''_{(can)}}{u_{(can)}}\right)h'-Gh=0. 
\end{equation}

\noindent With $\phi$ given by equation (\ref{phig}), one can note that $G$ is a small term, thus the solution is

\begin{equation}\label{h}
h=1+ \int \frac{1}{u_{(can)}}\left[\int G u^{2}_{(can)}d\eta\right]d\eta. 
\end{equation}

\noindent With $\Psi \simeq uf $, one can find 


\begin{equation}
\Psi \simeq u_{(can)}\left(1+ \int \frac{1}{u_{(can)}}\left[\int G u^{2}_{(can)}d\eta\right]d\eta\right)f.
\end{equation}

\noindent Note that $\Psi$ can be rewritten in an intuitive form by integrating $f$, which gives $f=(\rho+p)^{1/2}\exp{[-2\int \phi d\eta]}$, in such way 

\begin{equation}\label{psikp}
\Psi=\Psi_{(can)}he^{-2\int \phi d\eta},
\end{equation}

\noindent where $\Psi_{(can)}=(\rho+p)^{1/2}u_{(can)}$. This equation shows that the torsion acts like a damping in scalar perturbation, then, it predicts a red-tilted spectral index, $n_{s}$, in some $\alpha$ range of values, as it will be discussed below.\\ 

The power spectrum of $\Psi$ is given by $\mathcal{P}_{s}=|\Psi|^{2}k^{3}$,  and using ($\ref{h}$) we have:

\begin{equation}
\mathcal{P}_{s} \simeq \mathcal{P}_{s(can) }\left(1+ \int \frac{1}{u_{(can)}}\left[\int G u^{2}_{(can)}d\eta\right]d\eta\right)^{2}g^{2},
\end{equation}

\noindent with $\mathcal{P}_{s(can)}$ being the canonical power spectrum, and $g=\exp{[-2\int \phi d\eta]}$. One may look for the relation between the power spectrum and spectral index, which is given by

\begin{equation}
\mathcal{P}_{s} \propto k^{n_{s}-1}.
\end{equation}

Now, aiming on spectral index, we need to take the $\ln$ of the above equation:

\begin{eqnarray}
\ln {\mathcal{P}}_{s} \simeq \ln\mathcal{P}_{s (can)} &+& 2\ln \left(1+ \int \frac{1}{u_{(can)}}\left[\int G u^{2}_{(can)}d\eta\right]d\eta\right)\nonumber\\
&+&2\ln g.
\end{eqnarray}

\noindent The spectral index, in its turn, is given by

\begin{equation}
n_{s} - 1 = \frac{d \ln \mathcal{P}_{s}}{d \ln k}.
\end{equation}

Assuming that torsion does not affect the time of Hubble Horizon crossing, one have $d\ln k = \mathcal{H}^{-1}d\eta$, thus

\begin{eqnarray}
n_{s} &-& 1 = \frac{1}{\mathcal{H}}\left(\ln \mathcal{P}_{s( can)}\right)'+\frac{2}{\mathcal{H}}\left[\ln \left(1+ \int \frac{1}{u_{(can)}}\left[\int G u^{2}_{(can)}d\eta\right]d\eta\right)\right]'+\frac{(\ln g)'}{\mathcal{H}},
\end{eqnarray}

\noindent with $
(\ln g)' =-2(1+c^{2}_{s})\mathcal{\phi}$.

\noindent Although we not have a matter field rolling down into a bottom of a cosmological potential, the slow-roll condition $H'/H^{2}$ still hold. In this scenario, as pointed by Mukhanov \cite{Mukhanov:2005sc}, the time derivative of logarithmic quantity divide by $\mathcal{H}$ must be very small. Using (\ref{phig}, \ref{rhoinflac},  \ref{alphaomega}), and assuming the perturbation in the galactic scale, i.e


\begin{equation}
\frac{1}{\mathcal{H}}\left(\mathcal{P}^{2}_{s (can)}\right)' \simeq -3 \left(1+\frac{p}{\rho}\right)_{can},
\end{equation}

\noindent it is possible to find out the following spectral index  

\begin{equation}
n_{s}=1-3 \left(1+\frac{p}{\rho}\right)_{can}-\frac{8 \alpha ^2}{3-6 \alpha }.
\end{equation}

\noindent Assuming $-3 \left(1+p/\rho\right)_{can}\simeq 10^{-2}$, the above equation fix the $\alpha$ values in the range  $-0.157 \leqslant \alpha \leqslant 0.12$, leading to

\begin{equation}
0.92\lesssim n_{s}\lesssim 0.97.
\end{equation}

\noindent Interesting to notice that the interval of $\alpha$ is very small. Comparing with $\alpha$ previously fixed in \cite{Pereira:2019yhu} for recent acceleration, in the range $-0.01\leq \alpha \leq 0.81$, we found an intersection range of $\alpha = [-0.01,0.12]$, which could be used in a kind of unification model of inflation and accelerating phase.

\begin{figure}[htb]
\center{\includegraphics[width=10.0cm]
{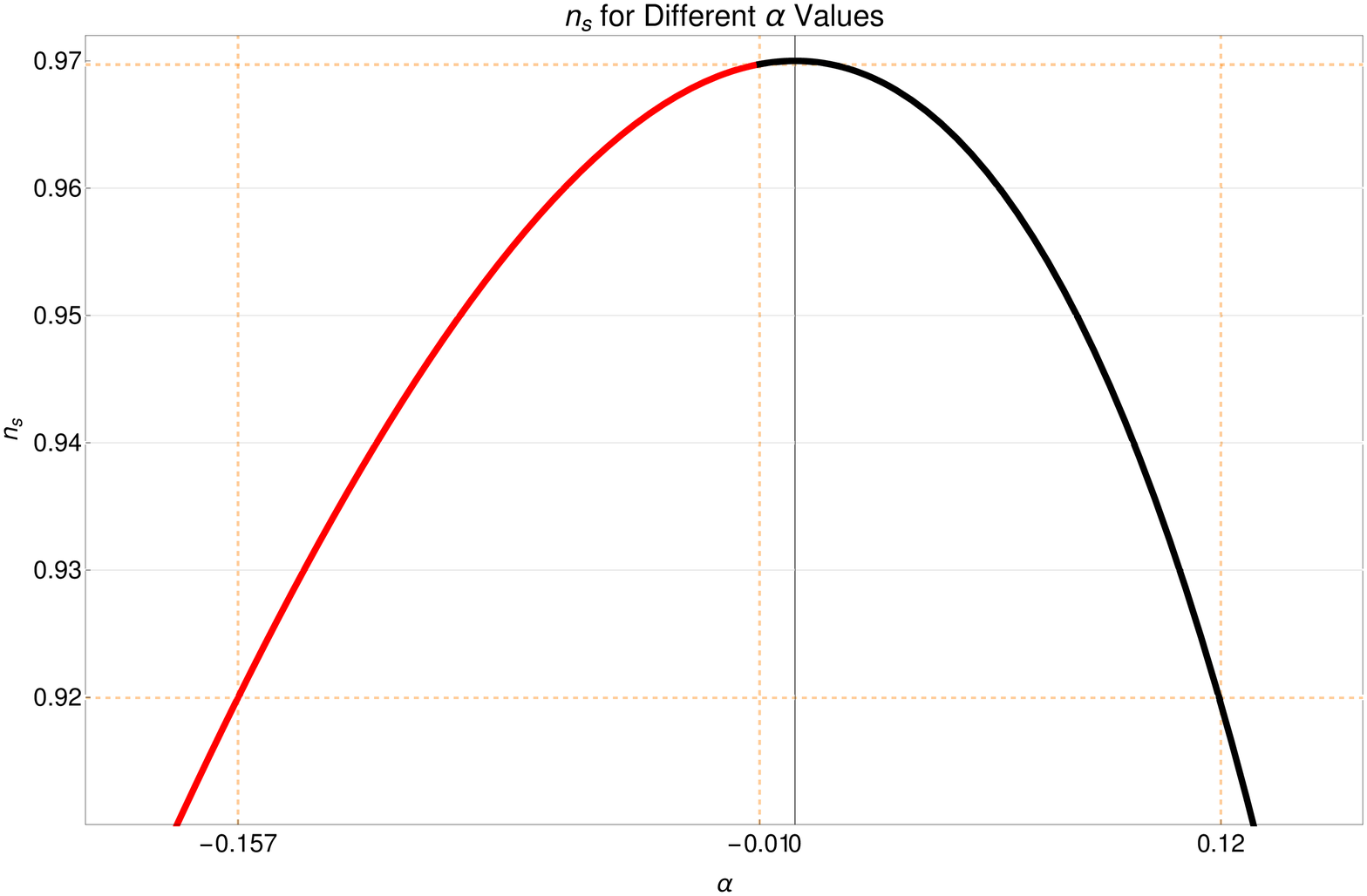}}
\vspace{0.1in}
\caption{\label{fig2} Variation of $n_{s}$ over the $\alpha$ range of values. Note that the spectated values of $n_{s}$ relies in the $\alpha$ interval $[-0.157,0.12]$, but $-0.157\leq\alpha<0.01$ (red part of the above curve) is out of the range previously assumed in \cite{Pereira:2019yhu}}.
\end{figure}

The Fig (\ref{fig2}) shows how $n_{s}$ changes with different $\alpha$ values inside the range $[-0.157,0.12]$, note that the black part of the curve is in agreement with the range previously assumed.

\subsection{Short-wavelength limit}

In the short-wavelength limit, the equation (\ref{mk2}) with torsion assumes the following form 

\begin{equation}
u'' + k^{2} u=0, 
\end{equation}

\noindent with the following solution

\begin{equation}\label{psikg}
\Psi=\Psi_{(can)}e^{-2\int\phi d\eta}.
\end{equation}

\noindent Thus, we recover the same equation for $n_{s}$ as the long-wave-length limit;

\begin{equation}
n_{s}=1-3 \left(1+\frac{p}{\rho}\right)_{can}-\frac{8 \alpha ^2}{3-6 \alpha },
\end{equation}

\noindent with $-0.157\leqslant \alpha \leqslant 0.12$.\\

In both cases, short and long wave-length regime, the torsion gives a small negative contribution to spectral index, as expected due to the ``damping" behaviour of $\phi$ in $\Psi$. Nevertheless, $\alpha$ could be measured. \\

The last, but not least, we have to find the tensor-to-scalar ratio, $r$. Note that the torsion does not change the metric, and $K^{i}{}_{jk}=0$, for $i,j,k$ being spatial index, then, the power spectrum assumes the canonical form 
\begin{equation}
\mathcal{P}_{T (can)} \simeq \frac{8}{\pi}\mathcal{H}^{2}.
\end{equation}

Owing to the small torsion contribution to spectral index, the tensor-to-scalar ratio remains unchanged  

\begin{equation}
r = 2 \frac{\mathcal{P}_{T}}{\mathcal{P}_{s}}<0.11.
\end{equation}

As it shown, the present model is compatible with CMB anisotropy. As $\alpha \rightarrow 0$, the canonical perturbations are recovered.

It is worth noting that the range of values we obtained for $\alpha$, both in the description of the inflationary era and for the analysis of perturbation theory with the presence of torsion coming from $\phi$, are in accordance with a sub-range present in the numerical study of the evolution of the late universe \cite{Pereira:2019yhu} for the description of its acceleration. In addition, once the $\alpha$ range has been fixed in this section, using the intersection values for the different phases of the universe, i.e. to the early and late universe, the general aspect of the behavior of torsion function, equation \eqref{phig}, is interesting for future studies.

\section{Conclusion}

In this paper, our aim has been to study a cosmological model with scalar torsion function $\phi$, which act like a constraint between the matter and geometry of the space-time. 

From the section \eqref{sec:inflation}, we analyze the analytical solutions of the evolution of the scale factor of a FLRW universe in the inflationary period. By the behavior of the temporal function $\phi$, \eqref{phig}, we describe the Hubble constant using the first Friedmann equation, eq. \eqref{4}. And, with the continuity equation, eq. \eqref{6}, we analyze the co-relation between the $\omega$ parameter, associates the density of matter and pressure contained in the universe, with the ${\alpha}$ parameter, which is responsible for dictating the influence of the torsion generated by the $\phi$ function. As a consequence, we obtain the description for the scale factor, revealing a good approximation with the fast growth rate of the universe in the era of inflation. In Section \eqref{sec:constraint}, we have shown that, in the period of low energy after the inflation, the behavior of the density of matter and the scale factor \eqref{rhocosmoatual} is very similar to $\Lambda$CDM, differing by a small contribution of $\alpha$, in the limit with ${\alpha \rightarrow 0}$, the canonical behavior of $\rho(t)$ and $a(t)$ is recovered.    

In section (\ref{sctep}), we have calculated the power spectrum and spectral index for a metric perturbation. As it shown, the cosmological perturbation, stretched out by inflation in the presence of the torsion function $\phi$, can be correctly described. In the equations (\ref{psikp}) and (\ref{psikg}), the torsion acts like a damping in the scalar perturbation $\Psi$, thus it represents a small negative correction in spectral index, restricting the range of values of $\alpha$ parameter in $[-0.1,0.12]$ interval. Once ${\delta T
^{i}{}_{j}}_{(T)}=0$, the equation (\ref{gw}) represents source-less gravitational waves from inflation, with no torsion contribution for it. Furthermore, the tensor-to-scalar ratio, $r$, still holding the same form, inasmuch as $\mathcal{P}_{T}$ does not change. It should be noted that, if $\alpha \rightarrow 0$, we recover the canonical equations over the entire model.

As a final remark, we would like to draw attention to fact that there is no re-heating phase after the inflation. In other words, this model shows an abrupt transition between inflation and matter dominant era, which should be carefully analysed in the future.

\vspace{0.5cm}

{\small {\bf Acknowledgements}}
{\small This study was financed in part by the Coordena\c{c}\~ao de Aperfei\c{c}oamento de Pessoal de N\'ivel Superior - Brasil (CAPES) - Finance Code 001. RdeCL would like to thank CAPES - Finance Code 001. SHP would like to thank CNPq - Conselho Nacional de Desenvolvimento Cient\'ifico e Tecnol\'ogico, Brazilian research agency, for financial support, grants number 303583/2018-5.}

\vspace{0.5cm}

\bibliographystyle{unsrt}  


\begin{thebibliography}{1}
\bibitem{Wald:1984rg}
R.~M.~Wald,
{\it General Relativity}, University of Chicago Press (1984).

\bibitem{Hehl:1976kj}
F.~W.~Hehl, P.~Von Der Heyde, G.~D.~Kerlick and J.~M.~Nester,
Rev. Mod. Phys. \textbf{48}, 393 (1976).

\bibitem{Hehl:1974cn}
F.~W.~Hehl, G.~D.~Kerlick and P.~Von Der Heyde,
Phys. Rev. D \textbf{10}, 1066 (1974).

\bibitem{Sabbata90}
V. de Sabbata, C. Sivaram, Astrophys. Space Sci. 165 (1990) 51.

\bibitem{Sabbata94}V. de Sabbata, C. Sivaram, Spin and Torsion in Gravitation, World Scientific,
1994.

  
\bibitem{Poplawski:2009su} 
  N.~J. Poplawski,
  Phys.\ Lett.\ B {\bf 690}, 73 (2010)
  Erratum: [Phys.\ Lett.\ B {\bf 727}, 575 (2013)];
  [arXiv:0910.1181 [gr-qc]].

\bibitem{Sciama:1964wt} 
  D.~W.~Sciama,
  Rev.\ Mod.\ Phys.\  {\bf 36}, 463 (1964)
  Erratum: [Rev.\ Mod.\ Phys.\  {\bf 36}, 1103 (1964)].
  
  \bibitem{Hehl:1971qi} 
  F.~W.~Hehl and B.~K.~Datta,
  J.\ Math.\ Phys.\  {\bf 12}, 1334 (1971).
  
  
\bibitem{Shapiro:2001rz} 
  I.~L.~Shapiro,
  Phys.\ Rept.\  {\bf 357}, 113 (2002);
  [hep-th/0103093].

\bibitem{tsamPRD} M. Tsamparlis, 
Phys. Rev. D {\bf 24}, 1451 (1981).

\bibitem{Pasmat2017}  K. Pasmatsiou, C. G. Tsagas, J. D. Barrow, Phys. Rev. D {\bf 95}, 104007 (2017).

\bibitem{Kranas2019}D. Kranas, C.G. Tsagas, J.D. Barrow, D. Iosifidis, Eur. Phys. J. C {\bf 79}, 341  (2019).


\bibitem{Barrow2019} J. D. Barrow, C. G. Tsagas, G. Fanaras, Eur. Phys. J. C {\bf 79}, 764 (2019).

\bibitem{Pereira:2019yhu}
S.~H.~Pereira, R.~d.~C.~Lima, J.~F.~Jesus and R.~F.~L.~Holanda,
Eur. Phys. J. C \textbf{79} (2019) no.11, 950
[arXiv:1906.07624 [gr-qc]].

\bibitem{Capozziello:2008kb} 
  S.~Capozziello, R.~Cianci, C.~Stornaiolo and S.~Vignolo,
  Phys.\ Scripta {\bf 78}, 065010 (2008);
  [arXiv:0810.2549 [gr-qc]]. 

\bibitem{Poplawski:2010kb}
  N.~J.~Poplaswski,
  Phys.\ Lett.\ B {\bf 694}, 181 (2010)
  Erratum: [Phys.\ Lett.\ B {\bf 701}, 672 (2011)];
  [arXiv:1007.0587 [astro-ph.CO]].

\bibitem{Olmo:2011uz} 
  G.~J.~Olmo,
  Int.\ J.\ Mod.\ Phys.\ D {\bf 20}, 413 (2011);
  [arXiv:1101.3864 [gr-qc]].

\bibitem{Jimenez:2015fva} 
  J.~Beltran Jimenez and T.~S.~Koivisto,
  Phys.\ Lett.\ B {\bf 756}, 400 (2016);
  [arXiv:1509.02476 [gr-qc]].  
  
\bibitem{Marques:2019ifg}
C.~M.~J.~Marques and C.~J.~A.~P.~Martins,
Phys. Dark Univ. \textbf{27} (2020), 100416
[arXiv:1911.08232 [astro-ph.CO]].

\bibitem{Bose:2020mdm}
A.~Bose and S.~Chakraborty,
Eur. Phys. J. C \textbf{80} (2020) no.3, 205;
[arXiv:2003.07226 [gr-qc]].

\bibitem{Bolejko:2020nbw}
K.~Bolejko, M.~Cinus and B.~F.~Roukema,
Phys. Rev. D \textbf{101} no.10, 104046 (2020);
[arXiv:2003.06528 [astro-ph.CO]].

\bibitem{Cruz:2020hkh}
M.~Cruz, F.~Izaurieta and S.~Lepe,
Eur. Phys. J. C \textbf{80} no.6, 559 (2020);
[arXiv:2005.04550 [gr-qc]].

\bibitem{Gonzalez-Espinoza:2019ajd} 
  M.~Gonzalez-Espinoza, G.~Otalora, N.~Videla and J.~Saavedra,
  JCAP {\bf 1908} (2019) 029.
  
\bibitem{Gasperini:1986mv}
M.~Gasperini,
``Spin Dominated Inflation in the Einstein-cartan Theory,''
Phys. Rev. Lett. \textbf{56}, 2873 (1986).
  
  
\bibitem{Valdivia:2017sat} 
  J.~Barrientos, F.~Cordonier-Tello, F.~Izaurieta, P.~Medina, D.~Narbona, E.~Rodr\'iguez and O.~Valdivia,
  Phys.\ Rev.\ D {\bf 96}, 084023 (2017).
  
  \bibitem{Izaurieta:2019dix} 
  F.~Izaurieta, E.~Rodr\'iguez and O.~Valdivia,
  Eur.\ Phys.\ J.\ C {\bf 79}, 337 (2019).
  
\bibitem{March:2011ry} 
  R.~March, G.~Bellettini, R.~Tauraso and S.~Dell'Agnello,
  Phys.\ Rev.\ D {\bf 83}, 104008 (2011);
  [arXiv:1101.2789 [gr-qc]].

\bibitem{Hehl:2013qga} 
  F.~W.~Hehl, Y.~N.~Obukhov and D.~Puetzfeld,
  Phys.\ Lett.\ A {\bf 377}, 1775 (2013);
  [arXiv:1304.2769 [gr-qc]].
  
\bibitem{Mukhanov:1990me}
V.~F.~Mukhanov, H.~A.~Feldman and R.~H.~Brandenberger,
Phys. Rept. \textbf{215}, 203-333 (1992). 

\bibitem{Basak:2011wp}
A.~Basak and J.~R.~Bhatt,
JCAP \textbf{06}, 011 (2011)
doi:10.1088/1475-7516/2011/06/011;
[arXiv:1104.4574 [astro-ph.CO]].  


\bibitem{Mukhanov:2005sc}
V.~Mukhanov, {\it Physical Foundations of Cosmology}, Cambridge University Press (2005). 
\end{thebibliography}

\end{document}